# Extending Machine Learning Based RF Coverage Predictions to 3D

Muyao Chen, Mathieu Châteauvert, Jonathan Ethier
Communications Research Centre Canada
Ottawa, Ontario, Canada
alice.chen@ised-isde.gc.ca, mathieu.chateauvert@ised-isde.gc.ca, jonathan.ethier@ised-isde.gc.ca

*Abstract*—This paper discusses recent advancements made in the fast prediction of signal power in mmWave communications environments. Using machine learning (ML) it is possible to train models that provide power estimates that are both accurate and with real-time simulation speeds. Work involving improved training data pre-processing as well as 3D predictions with arbitrary transmitter height is discussed.

## I. INTRODUCTION

The accurate prediction of communication system metrics is vital for the efficient and robust deployment of wireless networks. It is well established that deep learning techniques, including machine learning, can enhance the tools that operators and regulators use to analyze these deployments, via faster predictions or deeper insights into trends in existing data [1].

One area of interest is the prediction of radio frequency (RF) power in communications environments. This type of prediction provides access to quality-of-service metrics as well as insight into interference levels between adjacent deployments. It has been shown that convolutional neural networks (CNNs) can provide accurate power predictions when trained with simulation data from traditional RF power simulation tools [2]. CNN-based modeling work was done in [3] and [4] where ray-tracing simulation software [5,6] was used to generate the training data, forming the basis of prediction models. This offered more accurate simulations but presented additional challenges due to the complexity of the predicted power distributions. The work was shown to successfully provide accurate predictions relative to their ray-tracing tool counterparts, though with significant simulation speed improvements.

The work in this paper is a continuation of [3,4] with a new focus exploring more efficient pre-processing techniques for training data and constructing models with arbitrary transmitter height placements leading to full 3D prediction capabilities.

## II. RECENT PRE-PROCESSING TECHNIQUE ADVANCEMENT

Ray tracing simulation outputs can often be noisy. In order to address the noise issue, we apply an algorithm [7] that yields a locally time-averaged result of the rays impinging on each simulation point in the scene, resulting in physically based data smoothing. The small variations in the ray-tracing output power would increase the prediction error as the ML model cannot learn how to predict noisy variations. By decreasing the noise in the training and test sets, the mean absolute error (MAE) was reduced from 1.42 to 0.55 dB. These test sets have power maps with 1 meter resolution per pixel, 32x32 scene size, 28 GHz operating frequency, 720k training samples and 180k test samples. As a soundness check, simulations were performed in scenes with no buildings (i.e., empty space with terrain) and the MAE of predictions were as low as 0.13 dB, approaching zero error. This is as one would expect since ML ought to learn simple scenes lacking non-line-of-sight with ease. Similar improvements were observed for larger scenes.

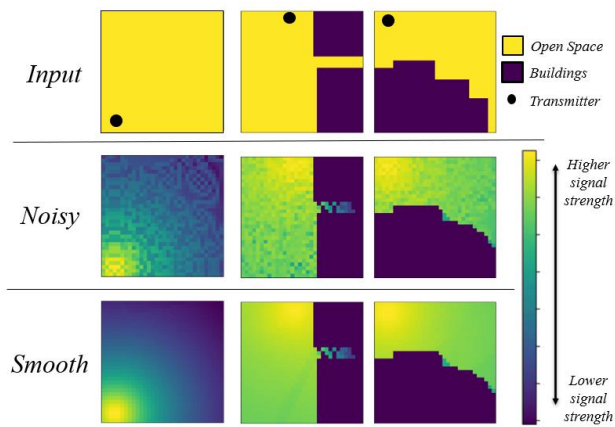

Fig. 1. Impact of the local time-averaging smoothing algorithm.

## III. EXTENDING THE MODEL TO 3D PREDICTIONS

This section is primarily focused on investigating the possibility of extending the current 2D RF simulation tool to 3D, motivated by the need to vary the transmitter height to achieve a more flexible prediction tool. In this study, an ML approach has been developed to optimize outdoor wireless coverage using pixel-wise regression models and CNNs. The 3D prediction model offers an advantage over 2D predictions by allowing the transmitter position to vary in 3D (as opposed to just 2D) while still maintaining a fast prediction rate.

### A. Data Acquisition and Preparation

To generate 3D volumetric inputs, multiple 2D planes at different elevations were concatenated. In order to limit the amount of data to be processed, we restricted the prediction area to "street-level" (building heights from 0 to 20 meters) to allow simulations in locations with higher levels of human activity while simultaneously reducing the number of layers used to describe building volumes. Transmitters were set between 12 to 20 meters to increase cell radius. The range of transmitter and receiver heights is arbitrary and can be modified as required based on deployment use cases.

*B. Power Simulation Scenarios*

Following pre-processing of the data, different methods were implemented to construct the models capable of 3D predictions. Fig. 2 illustrates the ML structures. We designed two different model types to simulate power in an urban scene:

**3D-to-2D prediction**: this algorithm generates a power map (or heat map) for a given 2D layer specified by the user based on 3D scenes with arbitrary transmitter locations.

**3D-to-3D prediction**: this approach takes 3D buildings with arbitrary transmitter parameters as inputs (same as 3D-to-2D) but in this case their corresponding 3D power volumes (or heat volumes) are simulated.

In both model types, the image size is 128x128 and the number of input layers (channels) is 5 with a physical separation of 4 meters. In the 3D-to-3D model, the number of output layers matches the number of inputs layers with the same layer separation.

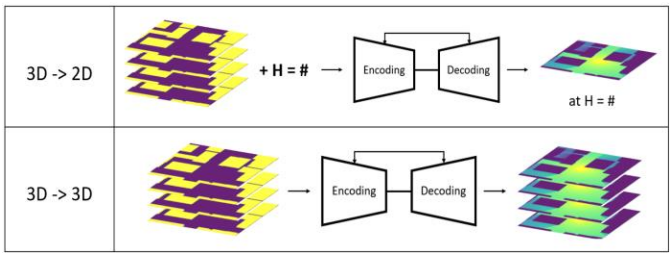

Fig. 2. Prediction architectures with various types of output structures.

*C. Deep Learning Model Training and Test*

In this study, a deep UNet architecture [2] is applied to construct the CNNs. Specifically, the model architecture is an encoder-decoder framework with skip connections. These skip connections were shown in [2-4] to provide enhanced accuracy over CNNs that follow the generic encoder-decoder topology.

We divided the urban scenes into two disjointed regions, one serving as training sets for developing the model, and the other being used as validation sets for testing the model's performance. MAE was used as an error metric to evaluate the performance of the model; it measures the average difference between simulation [6] and predictions without considering the predictions inside buildings.

In a 3D-to-2D approach, an MAE of 2.20 dB was obtained, with a simulation time of 0.015 seconds (s) for one specified layer and 0.074s for multiple (five) layers. As for the 3D-to-3D approach, an MAE of 2.49 dB was achieved, but the prediction time for multiple layers was 0.023s, which was three times faster than the 3D to 2D approach when generating multiple layers. Specifically, the 3D-to-3D architecture is ideal for scenarios where the complete 3D set of outputs is required, while the 3D-to-2D option is intended for faster simulation speed and is suitable for predicting a single layer.

The distribution of predicted power compared to simulated power is shown in Fig. 3. It is evident from the overlapping distributions that our modeling method is accurate. Some challenges remain for low power predictions (in the vicinity of –100 dBm) which will be addressed in future work.

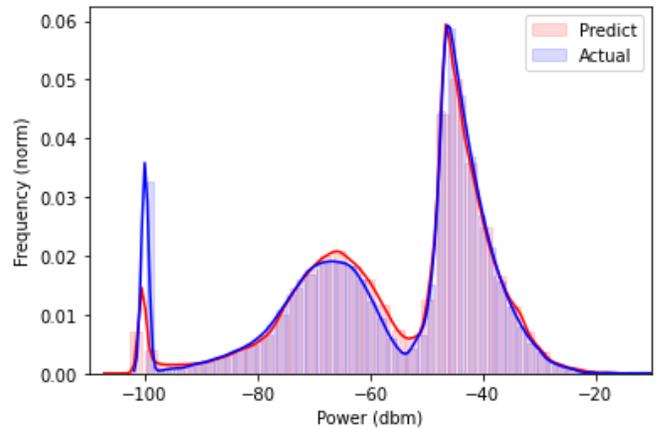

Fig. 3. Histogram of power simulations and model predictions provided by the 3D-to-3D model using the test set.

In addition, experiments were conducted to examine predictions with different frame sizes, e.g., as low as 64 x 64 and up to 256 x 256 pixels. It is feasible to construct models capable of real-time predictions using these larger power map sizes, but work still remains to improve the prediction accuracy.

## IV. CONCLUSIONS

This work explores recent advancements in ML-accelerated real-time RF power predictions. An efficient data pre-processing technique was discussed. Additionally, the models were shown to be successfully extended into 3D predictions, allowing one to predict RF power at arbitrary transmitter heights, including 3D building layouts as input and 3D power predictions as outputs.